# A Linear Time Algorithm for Solving #2SAT on Cactus Formulas

M. A. López, J. R. Marcial, G. De Ita, H. A. Montes-Venegas and R. Alejo

*Abstract*— An $O(n+m)$-time algorithm is presented for counting the number of models of a two Conjunctive Normal Form Formula $F$ that represents a Cactus graph, where $n$ is the number of variables and $m$ is the number of clauses of $F$. Although, it was already known that this class of formulas could be computed in polynomial time, we compare our proposal algorithm with two state of the art implementations for the same problem, *sharpSAT* and *countAntom*. The results of the comparison show that our algorithm outperforms both implementations, and it can be considered as a base case for general counting of two Conjunctive Normal Formulas.

## I. INTRODUCCIÓN

EL PROBLEMA $SAT(F)$, donde $F$ es una fórmula Booleana, consiste en decidir si $F$ tiene un modelo, es decir, una asignación a las variables de $F$ tal que, al ser evaluada con respecto a la lógica proposicional, devuelve un valor verdadero. Si $F$ esta en dos Forma Normal Conjuntiva (2-FNC), entonces $SAT(F)$ se puede resolver en tiempo polinomial. Sin embargo, si $F$ esta en $k$-FNC, $k > 2$, entonces $SAT(F)$ es un problema NP-Completo. Por otro lado, existe el problema de conteo denotado como $\#SAT(F)$ que consiste en contar el número de modelos de $F$. A diferencia de $SAT(F)$ que se puede considerar como un problema de decisión, $\#SAT(F)$ es un problema de conteo. $\#SAT(F)$ pertenece a la clase $\#P$-Completo aún cuando $F$ este en 2-FNC, este último denotado como $\#2SAT$ [1].

Aunque el problema $\#2SAT$ es $\#P$-Completo, existen instancias que se pueden resolver en tiempo polinomial [2]. Por ejemplo, si la gráfica que representa la fórmula es acíclica, entonces $\#2SAT$ puede resolverse en tiempo polinomial. $\#2SAT$ es considerado un problema fundamental en el establecimiento de la frontera entre problemas de conteo intratables y aquellos que pueden resolverse eficientemente.

$\#SAT(F)$ puede reducirse a diferentes problemas en el área de razonamiento aproximado. Por ejemplo, cuando se quiere estimar el grado de creencia, en la generación de explicaciones para consultas en bases de datos lógicas, en la inferencia Bayesiana y en el mantenimiento de sistemas de razonamiento [3][4][5]. Los anteriores problemas provienen de aplicaciones de la Inteligencia Artificial, tales como planeación, sistemas expertos, razonamiento automático, etc.

Actualmente, los algoritmos utilizados para $\#SAT(F)$ para cualquier fórmula $F$ en 2-FNC, descomponen $F$ en subfórmulas hasta obtener casos base en los que se puede contar modelos de forma eficiente. El algoritmo de menor orden de complejidad conocido hasta el momento fue desarrollado por Wahlström [3], éste es de $O(1.2377^n)$, donde $n$ representa el número de variables de la fórmula. El algoritmo de Wahlström utiliza como criterio de elección de una variable, el número de veces que aparece en la fórmula (sea la variable o su negación). Los dos criterios de paro del algoritmo son cuando $F = \emptyset$ o cuando $\emptyset \in F$.

Por otro lado, están las implementaciones para $\#SAT$ en donde el objetivo es buscar estrategias que permitan resolver el problema de forma eficiente para instancias en donde el número de variables es considerable. Las herramientas que a la fecha se consideran las más eficientes son *relsat* [6] y *sharpSAT* [7], ambas bajo el paradigma secuencial. También se han implementado métodos paralelos, como: *countAntom* [8].

*relsat* esta basado en el algoritmo DPLL (Davis-Putnam algorithm) cuya ventaja es poder procesar rápidamente fórmulas disjuntas, es decir, cuyos conjuntos de variables no se intersectan. La herramienta *sharpSAT* es una mejora de *relsat* en donde se elige una literal heurísticamente. Asimismo, utiliza descomposición y *cache* de componentes para agilizar el cálculo. La herramienta *countAntom* es una implementación paralela basada en *sharpSAT* cuyos resultados muestran que se mejora el tiempo en las instancias de prueba con respecto a *sharpSAT*.

En este artículo se presenta un algoritmo de compeljidad lineal en tiempo para el conteo de modelos de fórmulas en 2-FNC y cuya gráfica de restricciones sea tipo cactus. Los experimentos presentados muestran que nuestra propuesta mejora sustancialmente el tiempo para realizar el conteo de modelos con respecto a las herramientas de software actuales; *relsat, sharpSAT* y *countAntom*, por lo que nuestro algoritmo puede utilizarse como caso base en los algoritmos de conteo de modelos basados en descomposición de fórmulas en FNC.

## II. PRELIMINARES

Sea $X = \{x_1, \ldots, x_n\}$ un conjunto de $n$ variables Booleanas. Una literal es una variable $x_i$ ($x_i^1$) o la variable negada $\neg x_i$ ($x_i^0$). Una cláusula es una disyunción de literales distintas. Una fórmula Booleana $F$ en forma normal conjuntiva (FNC) es una conjunción de cláusulas.

Sea $v(Y)$ el conjunto de variables involucradas en el

M. A. López, Universidad Autónoma del Estado de México, México, mlopezm158@alumno.uaemex.mx

J. R. Marcial, Universidad Autónoma del Estado de México, México, jrmarcialr@uaemex.mx

G. De Ita, Benemérita Universidad Autónoma de Puebla, México, deita@cs.buap.mx

H. A. Montes-Venegas, Universidad Autónoma del Estado de México, México, hamontesv@uaemex.mx

objeto $Y$, dónde $Y$ puede ser una literal, una cláusula o una fórmula Booleana. Por ejemplo, para la cláusula $c = \{x_1 \vee \neg x_2\}$, $v(c) = \{x_1, x_2\}$.

Una asignación $s$ para $F$ es una función Booleana $s: v(F) \rightarrow \{0,1\}$. Una asignación puede también considerarse como un conjunto de pares de literales no complementario. Si $x^e \in s$ $e \in \{0,1\}$, siendo $s$ una asignación, entonces $s$ convierte a $x^e$ en verdadero y a $x^{1-e}$ en falso. Considerando una cláusula $c$ y una asignación $s$ como un conjunto de literales, se dice que $c$ se satisface por $s$ sí y solo si $c \cap s \neq \emptyset$ y si para toda $x^e \in c$, si $x^{1-e} \in s$, entonces $s$ falsifica a $c$.

Sea $F$ una fórmula Booleana en FNC, se dice que $F$ se satisface por la asignación $s$ si cada cláusula en $F$ se satisface por $s$. Por otro lado, se dice que $F$ se contradice por $s$ si al menos una cláusula de $F$ se falsifica por $s$. Un modelo de $F$ es una asignación para $v(F)$ tal que satisface $F$.

Dada una fórmula $F$ en FNC, *SAT* consiste en determinar si $F$ tiene un modelo, mientras que *#SAT* consiste en contar el número de modelos que tiene $F$ sobre $v(F)$. Por otro lado, *#2SAT* denota *#SAT* para fórmulas en 2-FNC.

II.I. La gráfica restringida de una 2-FNC.

Existen algunas representaciones gráficas de una Forma Normal Conjuntiva, en este caso, se utilizará la gráfica primal signada (gráfica restringida) [9].

Sea $F$ una 2-FNC, su gráfica restringida se denota por $G_F = (V(F), E(F))$ donde los vértices de la gráfica son las variables $V(F) = v(F)$, y las cláusulas las aristas $E(F) = \{\{v(x), v(y)\}: \{x \vee y\} \in F\}$, esto es, para cada cláusula $\{x \vee y\}$ en $F$ existe una arista $\{v(x), v(y)\} \in E(F)$. Para $x \in V(F)$, $\delta(x)$ denota su grado, es decir, el número de aristas incidentes en $x$. Cada arista $c = \{v(x), v(y)\} \in E(F)$ se asocia con un par $(s_1, s_2)$ de signos, que se asignan como etiquetas de la arista que conecta las variables de la cláusula en la gráfica. Los signos $s_1$ y $s_2$ pertenecen a las variables $x$ y $y$ respectivamente. Por ejemplo, la cláusula $(x^0 \vee y^1)$ determina la arista con etiqueta "$x \xrightarrow{-+} y$", que es equivalente a la arista "$y \xrightarrow{+-} x$".

Sea $S = \{+, -\}$ un conjunto de signos. Una gráfica con aristas etiquetadas en $S$ es el par $(G, \psi)$, dónde $G = (V, E)$ es una gráfica restringida, y $\psi$ es una función con dominio $E$ y rango $S$. $\psi(e)$ se denomina a la etiqueta de la arista $e \in E$. Sea $G = (V, E, \psi)$ una gráfica restringida con aristas etiquetadas en $S \times S$ y $x$ y $y$ vértices en $V$, si $e = \{x, y\}$ es una arista y $\psi(e) = (s, s')$, entonces $s(s')$ es el signo adyacente a $x(y)$.

Note que la gráfica restringida $G_F = (V, E, \psi)$ de una 2-FC $F$ puede contener aristas paralelas. Nosotros consideraremos gráficas simples (sin aristas paralelas), ya que éstas últimas pueden ser pre-procesadas, tal y como se presenta en [4], sin que este pre-procesamiento modifique la complejidad en tiempo de nuestra propuesta algorítmica.

Sea $G$ una gráfica conectada de $n$ vértices, un árbol de expansión de $G$ es un subconjunto de $n - 1$ aristas tal que forman un árbol de $G$. Se denomina coárbol al subconjunto de aristas que son el complemento de un árbol, cada una de estas aristas forman un ciclo en la gráfica.

Una gráfica cactus es una gráfica $G = (V_G, E_G)$ dónde:

- Cada arista de $E_G$ pertenece a lo más a un ciclo.
- Cualquier par de ciclos comparten a lo más un vértice.

En este artículo, para encontrar el árbol de expansión y el coárbol se utiliza el método de búsqueda primero en profundidad [10] que permite construir la tupla (árbol, coárbol) de una gráfica.

III. ALGORITMO

En esta sección se presenta el algoritmo utilizado para el conteo de modelos en gráficas cactus en tiempo lineal. El método principal consiste de 3 pasos: construcción de una tabla en donde se almacenan las cláusulas, construcción de un árbol de expansión en donde se marcan los ciclos de la gráfica y finalmente, se realiza el conteo de modelos sobre el árbol de expansión. El algoritmo I muestra las entradas y salidas de cada paso.

ALGORITMO I
CONTEO DE MODELOS EN GRÁFICAS CACTUS

**Entrada:** Fórmula $F$
$t = $ construcción_Tabla(F, nF); //nF = # variables de F
$arbol = $ crear_Arbol(t);
$\{(\alpha_1, \beta_1), (\alpha_2, \beta_2)\} = $ cuenta(arbol);
return $\alpha_1 + \beta_1$;
**Salida:** Número de modelos de $F$

III.I. Construcción de la Tabla de Cláusulas.

Ya que la lectura de cláusulas se realiza desde un archivo en formato DIMACS (http://logic.pdmi.ras.ru/~basolver/dimacs.html), se utilizan dos tablas dinámicas para almacenar cada cláusula leída. El índice del renglón $h$ de cada tabla representa a la variable $x_h$. Por cada cláusula $(x_i, x_j)$, se agrega la entrada $x_j$ al renglón $i$ en la tabla uno, y la entrada $x_i$ al renglón $j$ en la tabla dos. La duplicidad de cláusulas permite agilizar las búsquedas durante la construcción del árbol de expansión. Adicionalmente, las cláusulas se insertan considerando el índice menor de sus dos variables. El algoritmo II muestra la construcción de la tabla.

III.II. Creación del árbol.

El par (árbol, coárbol) se crea a partir de las tablas que almacenan cláusulas. El algoritmo utilizado para la construcción del árbol es *primero en profundida* (*depth first search*, por sus siglas en inglés) [10]. Cada vez que se lee una cláusula, se revisa si ambos vértices están ya en el árbol, si es el caso, se marca el camino entre ambos vértices para denotar que se tiene una arista del coarbol, es decir, se encontró un ciclo.

El algoritmo III muestra la forma en que se recorre la tabla para insertar sus elementos en el árbol, mientras que el algoritmo IV presenta la forma de insertar una variable en el árbol considerando si ésta aparece de forma positiva o negativa en la cláusula de la que proviene.

ALGORITMO II
CONSTRUCCIÓN DE LA TABLA DE CLÁUSULAS

**Entrada:** Fórmula $F$, # variables de F $nF$
vector $t_1$ de tamaño $nF + 1$;
vector $t_2$ de tamaño $nF + 1$;
**para cada** *clausula* $C = \{v_1, v_2\}$ *de F* **hacer**
  **si** $v_1 \leq v_2$ **entonces**
    agregar $\{v_1, v_2\}$ a $t1[v_1]$;
    agregar $\{v_1, v_2\}$ a $t2[v_2]$;
  **en otro caso**
    agregar $\{v_2, v_1\}$ a $t1[v_2]$;
    agregar $\{v_2, v_1\}$ a $t2[v_1]$;
  **fin**
**fin**
return t;
**Salida:** Tabla de cláusulas t

Como se puede apreciar en el algoritmo IV, se tiene la condición para conocer si la gráfica de entrada es cactus, lo anterior con la finalidad de hacer una comparación equitativa con las herramientas *sharpSAT* y *countAntom*, ya que en ellas no se conoce de antemano si la gráfica de entrada es cactus.

ALGORITMO III
CONSTRUCCIÓN DEL ÁRBOL UTILIZANDO BÚSQUEDA PRIMERO EN PROFUNDIDAD

**Entrada:** Tablas de cláusulas $t_1, t_2$
$i = 1, T = NULL$;
**mientras** $t_1[i] == NULL$ and $t_2[i] == NULL$ **hace**
  i++
**fin**
**mientras** *true* **hacer**
  **si** $t_1[i] \neq NULL$ or $t_2[i] \neq NULL$ **entonces**
    **si** $t_1[i] \neq NULL$ **entonces**
      $c = t_1[i][0]$;
    **en otro caso**
      $c = t_2[i][0]$;
    **fin**
    i = agregar(c,T);
  **en otro caso**
    **si** $i$ *tiene padre* **entonces**
      $i$ = padre de $i$ en T;
    **en otro caso**
      return T;
    **fin**
  **fin**
**fin**
**Salida:** Un árbol T

III.III. Conteo de modelos.

Una vez que se tiene el árbol de la gráfica cactus con los caminos de los ciclos marcados (coárbol), se realiza el conteo de los modelos mediante un recorrido en postorden del árbol. A cada nodo $x_i$ del árbol se le asigna un par $(\alpha_i, \beta_i)$ donde $\alpha_i$ denota el número de modelos en donde la variable $x_i$ toma un valor verdadero y $\beta_i$ el número de modelos en donde $x_i$ toma un valor falso. Los modelos contabilizados por cada par serán relativos a la subfórmula hasta ese momento calculada.

Si el nodo del árbol esta marcado como de apertura de un ciclo, se genera un segundo par $(\alpha_{2_i}, \beta_{2_i})$ que estará activo hasta que el ciclo se cierre. El número de modelos sobre el nodo de cierre de ciclo, se calcula como: $(\alpha_i, \beta_i) - (\alpha_{2_i}, \beta_{2_i})$, donde la resta se realiza a pares.

ALGORITMO IV
AGREGAR VÉRTICES AL ÁRBOL Y DETERMINAR SI LA GRÁFICA ES CACTUS.

**Entrada:** Cláusula c=$v_1, v_2$, árbol T
Si $v_1$ no está en T, agregar $v_1$ como la raíz de T;
**si** $v_1 == v_2$ **entonces**
  si los signos de $v_1$ y $v_2$ son iguales marcar $v_1$ como unitario;
  si no es unitario ignorar la cláusula;
  return $v_1$;
**fin**
**si** $v_2$ *no está en T* **entonces**
  agregar $v_2$ como nodo hijo de $v_1$ return $v_2$;
**en otro caso**
  dibujar o marcar el camino de $v_1$ a $v_2$;
  **si** *si el camino ya fue marcado* **entonces**
    la gráfica no es cactus, termina el algoritmo;
  **en otro caso**
    marcar dependiendo del signo, el inicio (el nodo que se encuentre en un nivel más bajo del árbol) y cierre (la última arista marcada en el camino hacia el nodo final) del ciclo para el conteo;
  **fin**
**fin**
**Salida:** entero i

Inicialmente, a los nodos hoja se les asigna el par (1,1) y si el nodo hoja es a la vez la apertura de un ciclo, se le asigna como segundo par (0,1). Durante el recorrido, el par de cada nodo interior se calcula utilizando la recurrencia (1) que se aplica de acuerdo a los signos de las variables del nodo actual y del nodo hijo (es decir, de la cláusula representada por la arista hijo-padre en el árbol).

$$(\alpha_i, \beta_i) = \begin{cases} (\beta_{i-1}, \alpha_{i-1} + \beta_{i-1}) \; if \; (\epsilon_i, \delta_i) = (-,-) \\ (\alpha_{i-1} + \beta_{i-1}, \beta_{i-1}) \; if \; (\epsilon_i, \delta_i) = (-,+) \\ (\alpha_{i-1}, \alpha_{i-1} + \beta_{i-1}) \; if \; (\epsilon_i, \delta_i) = (+,-) \\ (\alpha_{i-1} + \beta_{i-1}, \alpha_{i-1}) \; if \; (\epsilon_i, \delta_i) = (+,+) \end{cases} \quad (1)$$

Si un nodo interior es la apertura de un ciclo, su segundo par se calcula dependiendo de su signo, la recurrencia (2) establece el valor para el segundo par

$$(\alpha_{2_i}, \beta_{2_i}) = \begin{cases} (0, \beta_{2_i-1}) & \text{si } s = + \\ (\alpha_{2_i-1}, 0) & \text{si } s = - \end{cases} \quad (2)$$

Finalmente, si el nodo que se esta evaluando tiene mas de un hijo, el número de modelos se calcula mediante el producto de los pares de cada uno de ellos. El algoritmo V presenta el conteo de los modelos en el árbol.

Más detalles del método de conteo se puede consultar en [11], en donde se muestra cómo contar el número de modelos para árboles y ciclos por separado. En este mismo artículo se puede ver la demostración de la validez del método. Finalmente, el total de modelos se obtiene cuando se llega al nodo raíz $x_r$ del árbol, y se deriva de la suma de sus dos componentes: $\alpha_r$ y $\beta_r$.

ALGORITMO V
CONTEO DE MODELOS EN EL ÁRBOL QUE REPRESENTA LA GRÁFICA CACTUS.

```
Entrada: árbol T
para cada  D hijo de T hacer
    {(α_{D_1}, β_{D_1}), (α_{D_2}, β_{D_2})} = cuenta(D);
fin
si T es hoja entonces
    (α_{T_1}, β_{T_1}) = (1, 1);
    (α_{T_2}, β_{T_2}) = (1, 1);
    aplicar (2) si es necesario;
    return {(α_{T_1}, β_{T_1}), (α_{T_2}, β_{T_2})};
en otro caso
    (α_r, β_r) = (1, 1);
    para cada  D hijo de T hacer
        aplicar (1) para {(α_{D_1}, β_{D_1}), (α_{D_2}, β_{D_2})};
        si la arista entre D y su padre T
        está marcada como cierre de ciclo entonces
            aplicar (2) para (α_{D_2}, β_{D_2});
            aplicar (3) (α_i, β_i) = (α_{D_1}, β_{D_1}) - (α_{D_2}, β_{D_2})
        fin
        (α_r, β_r) = (α_r, β_r) * (α_i, β_i);
    fin
    return {(α_r, β_r), (α_r, β_r)};
fin
Salida: {(α_1, β_1), (α_2, β_2)}
```

## IV. RESULTADOS

En esta sección se muestran los resultados de comparar la propuesta aquí presentada contra *sharpSAT* y *countAntom*, las dos herramientas más eficientes hasta el momento reportadas en la literarura para resolver #SAT. Se realizaron dos tipos de pruebas, la primera consistió en generar de manera aleatoria 22 gráficas cactus que tienen entre 9,000 y 240,000 vértices y entre 12,500 y 320,000 aristas. La segunda prueba consistió en generar gráficas cactus con el número máximo de ciclos que puede tener este tipo de gráficas, este es el peor de los casos en donde por cada tres vértices se generó un ciclo. Ya que *sharpSAT* y *countAntom* utilizan la estructura de la fórmula para realizar el conteo, éste segundo tipo de prueba generan los mejores casos para estas herramientas y para este tipo de gráficas.

Todas las pruebas se realizaron en una Computadora con procesador de dos núcleos a una velocidad de 2.4 Mhz con 8 GB en RAM y sistema operativo Ubuntu, ya que éste es el sistema operativo en donde se provee el código fuente para poder compilar y ejecutar ambas herramientas.

La Tabla 1 muestra los resultados de realizar el primer tipo de pruebas. Como se puede observar, nuestra propuesta obtiene mejores tiempos de ejecución en los 22 casos considerados, con respecto a las dos herramientas antes mencionadas. El tiempo de ejecución máximo para el conteo de modelos fue de 25 minutos. También se puede observar que *countAntom* no fue capaz de producir una respuesta para 17 de los 22 casos.

TABLA I
TIEMPO DE EJECUCIÓN DEL CONTEO DE MODELOS PARA 22 GRÁFICAS CACTUS. TIEMPO REPORTADO EN SEGUNDOS

|    | Vértices | Aristas | Ésta Propuesta | sharpSAT | countAntom |
|----|----------|---------|----------------|----------|------------|
| 1  | 9382     | 12508   | 0.076          | 0.213    | 11.297     |
| 2  | 19999    | 26664   | 0.056          | 0.490    | 851.425    |
| 3  | 10000    | 13332   | 0.060          | 0.231    | 427.430    |
| 4  | 16000    | 21332   | 0.065          | 0.380    | 197.042    |
| 5  | 25999    | 34664   | 0.071          | 0.700    | 1334.007   |
| 6  | 30001    | 40000   | 0.078          | 0.880    | ---------- |
| 7  | 36001    | 48000   | 0.098          | 1.1      | ---------- |
| 8  | 40000    | 53332   | 0.106          | 1.250    | ---------- |
| 9  | 46000    | 61332   | 0.118          | 1.509    | ---------- |
| 10 | 49999    | 66664   | 0.127          | 1.725    | ---------- |
| 11 | 55999    | 74664   | 0.144          | 2.016    | ---------- |
| 12 | 60001    | 80000   | 0.154          | 2.186    | ---------- |
| 13 | 66001    | 88000   | 0.167          | 2.540    | ---------- |
| 14 | 79999    | 106664  | 0.213          | 3.419    | ---------- |
| 15 | 100000   | 133332  | 0.260          | 4.828    | ---------- |
| 16 | 120001   | 160000  | 0.310          | 6.418    | ---------- |
| 17 | 139999   | 186664  | 0.361          | 8.354    | ---------- |
| 18 | 160000   | 213332  | 0.409          | 10.420   | ---------- |
| 19 | 180001   | 240000  | 0.457          | 12.734   | ---------- |
| 20 | 199999   | 266664  | 0.510          | 15.354   | ---------- |
| 21 | 220000   | 293332  | 0.558          | 18.293   | ---------- |
| 22 | 240001   | 320000  | 0.609          | 21.353   | ---------- |

La Gráfica 1 muestra el crecimiento en el tiempo de ejecución, tanto para *sharpSAT* como para nuestra propuesta. En la gráfica no se incluyen los tiempos de *countAntom*, ya que a partir de la sexta entrada, esta herramienta tardó mas de 25 minutos en dar una respuesta, por lo que se considera que sus tiempos de respuesta no son competitivos con respecto a los otros dos programas. Como se puede observar en la Gráfica 1, nuestra propuesta obtiene mejores tiempos de ejecución en todos los casos que los obtenidos con *sharpSAT*.

GRÁFICA I
CRECIMIENTO DEL TIEMPO DE EJECUCIÓN PARA 22 GRÁFICAS
CACTUS. EL TIEMPO REPORTADO ESTA EN SEGUNDOS

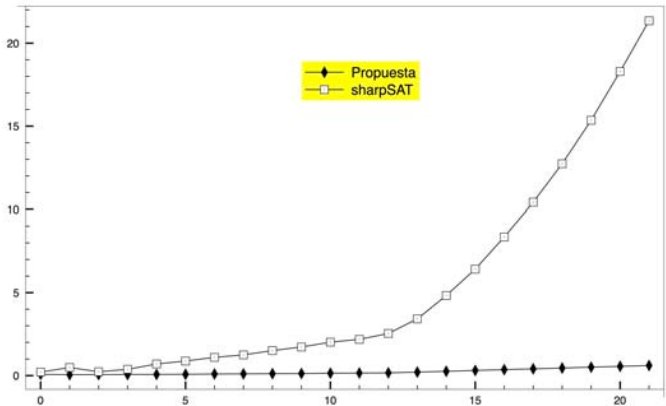

Para realizar pruebas considerando los peores casos para nuestra propuesta, se generaron gráficas cactus con el máximo número de ciclos posibles, es decir, se formó un ciclo por cada tres vértices. La diferencia principal con este tipo de gráficas es que el árbol generado tiene un tercio de nodos de inicio y un tercio de nodos de cierre de ciclos lo cual significa que se tienen dos variables de conteo en todo el recorrido del árbol.

La Tabla II muestra los resultados del tiempo de ejecución para el conteo de modelos sobre este tipo de gráficas. Como se puede observar en los 7 casos, nuestra propuesta obtiene menores tiempos de ejecución que el de las otras dos herramientas. Similar al primer caso, *counAntom* dio un resultado para los primero 5 casos, sin embargo, para los dos últimos no fue capaz de producir un resultado en los 5 minutos de ejecución que se dieron como máximo em esta segunda prueba.

TABLA II
TIEMPO DE EJECUCIÓN DEL CONTEO DE MODELOS EN 7
GRÁFICAS CACTUS CON EL MÁXIMO NÚMERO DE CICLOS.
TIEMPO REPORTADO EN SEGUNDOS

|   | Vértices | Aristas | Ésta Propuesta | sharpSAT | countAntom |
|---|----------|---------|----------------|----------|------------|
| 1 | 9999     | 14997   | 0.028          | 0.240    | 1.637      |
| 2 | 15999    | 23997   | 0.05           | 0.399    | 8.562      |
| 3 | 19999    | 29997   | 0.062          | 0.515    | 8.960      |
| 4 | 39999    | 59997   | 0.115          | 1.312    | 50.023     |
| 5 | 79999    | 119997  | 0.225          | 3.572    | 239.328    |
| 6 | 159999   | 239997  | 0.446          | 11.256   | ---------- |
| 7 | 239999   | 359997  | 0.669          | 23.413   | ---------- |

La Gráfica II muestra el crecimiento de los tiempos de ejecución del conteo de modelos para *sharpSAT* y para nuestra propuesta. Al igual que en la Gráfica I, no se muestran los resultados para *countAntom*, ya que sus tiempos reportados no son competitivos con respecto a los reportados para los otros dos programas. Similar a los resultados reportados para la primera prueba, nuestra propuesta mejora significativamente el tiempo de ejecución con respecto al de *sharpSAT*.

## IV. COMPLEJIDAD EN TIEMPO DEL ALGORITMO

El algoritmo de esta propuesta consta de tres partes, crear tabla, crear árbol y conteo de modelos en el árbol. Para la creación de la tabla es suficiente con recorrer las cláusulas de la fórmula de entrada por lo que la complejidad de este paso es de orden $O(m)$. Mientras que para creación del árbol, se recorre cada entrada en la tabla, lo que requiere también del orden $O(m)$ operaciones básicas.

Finalmente, el recorrido del árbol es en postorden, y mientras se visitan los nodos del árbol, se va también visitando cada una de las aristas (cláusulas) de la gráfica de restricciones, al mismo tiempo que se van aplicando las recurrencias (1) y (2). Este último proceso nos genera del orden de $O(n + m)$ operaciones básicas. Así, la complejidad en tiempo de nuestra propuesta algorítmica es lineal y de orden $O(n + m)$.

GRÁFICA II
CRECIMIENTO DEL DEL TIEMPO DE EJECUCIÓN PARA 7
GRÁFICAS CACTUS QUE REPRESENTAN EL PEOR CASO PARA
NUESTRA PROPUESTA. TIEMPO REPORTADO EN SEGUNDOS

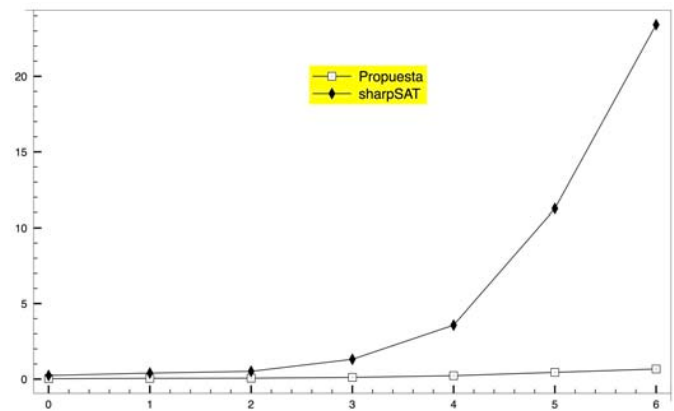

## V. CONCLUSIONES

En este artículo se presentó un algoritmo para el conteo de modelos de fórmulas Booleanas en 2-FNC y cuya gráfica de restricciones es tipo cactus. El algoritmo presentado tiene una complejidad en tiempo de orden $O(m + n)$, donde $m$ es el número de cláusulas y $n$ el número de variables de la fórmula de entrada. Nuestra propuesta detecta en tiempo lineal y al inicio de su procesamiento, si la fórmula de entrada es representada efectivamente por una gráfica de restricciones tipo cactus simple.

Las pruebas realizadas sobre dos diferentes clases de fórmulas muestran que la propuesta aquí presentada obtiene mejores tiempos de ejecución con respecto a las dos herramientas que se reportan en la literatura como las más eficientes; *sharpSAT* y *countAtom* para resolver este tipo de problemas de conteo. Por tanto, se considera que nuestro algoritmo puede incluirse como un caso base en el conteo de modelos de *sharpSAT* o de otras herramientas que realicen descomposición de la fórmula de entrada, típico de los métodos de ramificación y corte, hasta llegar a los casos bases.

Para el caso general #SAT, se podrían realizar descomposiciones sobre la fórmula de entrada, hasta generar gráficas cactus y entonces, procesar éstas últimas subfórmulas usando nuestro algoritmo, lo que creemos que puede impactar en la complejidad en tiempo de estos algoritmos de conteeo.

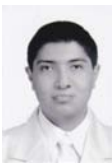

**Marco Antonio López Medina**, En el 2016 obtiene el título de Ingeniero en Computación en la Universidad Autónoma del Estado de México. Actualmente estudia la Maestría en Ciencias de la Ingeniería con línea de acentuación en Computación en la Universidad Autónoma del Estado de México.

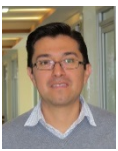

**José Raymundo Marcial-Romero**, Actualmente profesor investigador de tiempo completo en la Facultad de Ingeniería de la Universidad Autónoma del Estado de México. Miembro del Sistema Nacional de Investigadores del Consejo Nacional de Ciencia y Tecnología, Nivel 1. En el año 2000 obtuvo el título de Licenciatura en Ciencias de la Computación y en el año 2007 el grado de Doctor en Ciencias de la Computación por The University of Birmingham UK en The School of Computer Science.

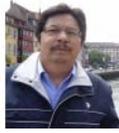

**Guillermo De Ita Luna**, Obtuvo su doctorado en Ingeniería Eléctrica por el CINVESTAV-IPN, México. Ha trabajado como desarrollador y consultor en sistemas de bases de datos y sistemas de información geográfica para diferentes empresas en México. Ha realizado estancias de investigación en la Universidad de Chicago, Texas A&M, INAOEP Puebla, en el instituto INRIA en Lille-1 y en la Fac. de Ing. de la UAEMEX. Actualmente, es profesor investigador de la Facultad de Ciencias de la Computación, BUAP, Puebla, México.

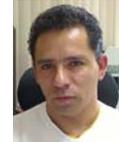

**Héctor Alejandro Montes-Venegas**, Es actualmente profesor investigador de tiempo completo en la Facultad de Ingeniería de la Universidad Autónoma del Estado de México. Es también ingeniero en Sistemas Computacionales por el Instituto Tecnológico de León y Maestro en Ciencias Computacionales por el Instituto Tecnológico y de Estudios Superiores de Monterrey.

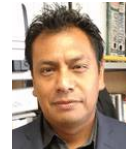

**Roberto Alejo Eleuterio** Doctor en Ciencias Computacionales, adscrito al Instituto de Estudio Superiores de Jocotitlán, Miembro del Sistema Nacional de Investigadores del CONACYT. Los intereses en investigación se centran en la aplicación de inteligencia artificial a la solución de problemas reales.